\documentclass[11pt,letterpaper]{article}
\usepackage[utf8]{inputenc}
\usepackage[T1]{fontenc}
\usepackage[margin=0.75in]{geometry}
\usepackage{mathptmx}
\usepackage{amsmath,amssymb,graphicx,needspace,titlesec}
\usepackage{xcolor}
\usepackage[colorlinks=true,urlcolor=blue,linkcolor=blue,citecolor=blue]{hyperref}
\hypersetup{pdftitle={Sensitive dependence of the motion of a legged robot on granular media},pdfauthor={Chen Li; Paul B. Umbanhowar; Haldun Komsuoglu; Daniel E. Koditschek; Daniel I. Goldman}}
\titleformat{\section}{\fontsize{13}{16}\selectfont\bfseries}{}{0pt}{}
\titlespacing*{\section}{0pt}{14pt}{7pt}
\renewenvironment{abstract}{\section*{Abstract}}{\par}
\let\originalthebibliography\thebibliography
\let\endoriginalthebibliography\endthebibliography
\renewenvironment{thebibliography}[1]{\originalthebibliography{#1}\fontsize{9.3}{11.625}\selectfont\setlength{\itemsep}{3pt}\setlength{\parskip}{0pt}}{\endoriginalthebibliography}
\makeatletter
\renewcommand{\@biblabel}[1]{#1.}
\makeatother
\newsavebox{\manuscriptimage}
\newcommand{\manuscriptfigure}[4]{%
  \par\begingroup
  \sbox{\manuscriptimage}{\includegraphics[#2]{#3}}%
  \Needspace{\dimexpr\ht\manuscriptimage+\dp\manuscriptimage+42pt\relax}%
  \noindent\makebox[\linewidth][c]{\hypertarget{fig#1}{}\usebox{\manuscriptimage}}\par
  \vspace{2pt}%
  \refstepcounter{figure}\label{figure#1}%
  {\fontsize{9.5}{12.35}\selectfont\noindent\textbf{Figure \thefigure.} #4\par}%
  \endgroup\vspace{8pt}%
}
\begin{document}
\fontsize{11.2}{15.23}\selectfont
{\centering\fontsize{18}{21.6}\selectfont\bfseries Sensitive dependence of the motion of a legged robot on granular media\par}
\vspace{8pt}
{\centering\fontsize{11}{15.4}\selectfont Chen Li$^{1}$, Paul B. Umbanhowar$^{2}$, Haldun Komsuoglu$^{3}$,\\ Daniel E. Koditschek$^{3}$, Daniel I. Goldman$^{1}$\par}
{\centering\fontsize{9.3}{12.1}\selectfont
$^{1}$School of Physics, Georgia Institute of Technology, Atlanta, GA 30332\\
$^{2}$Department of Mechanical Engineering, Northwestern University, Evanston, IL 60208\\
$^{3}$Department of Electrical and Systems Engineering, University of Pennsylvania, Philadelphia, PA 19104\par}
\vspace{4pt}
{\fontsize{9.5}{12.825}\selectfont
Published in Proceedings of the National Academy of Sciences 106(9), 3029--3034 (2009).\\
DOI: \href{https://doi.org/10.1073/pnas.0809095106}{10.1073/pnas.0809095106}\par}

\begin{abstract}
Legged locomotion on flowing ground ({\em e.g.}~granular media) is unlike locomotion on hard ground because feet experience both solid- and fluid-like forces during surface penetration. Recent bio-inspired legged robots display speed relative to body size on hard ground comparable to high performing organisms like cockroaches but suffer significant performance loss on flowing materials like sand. In laboratory experiments we study the performance (speed) of a small (2.3~kg) six-legged robot, SandBot, as it runs on a bed of granular media (1~mm poppy seeds). For an alternating tripod gait on the granular bed, standard gait control parameters achieve speeds at best two orders of magnitude smaller than the 2~body lengths/s ($\approx 60$~cm/s) for motion on hard ground. However, empirical adjustment of these control parameters away from the hard ground settings, restores good performance, yielding top speeds of 30~cm/s. Robot speed depends sensitively on the packing fraction $\phi$ and the limb frequency $\omega$, and a dramatic transition from rotary walking to slow swimming occurs when $\phi$ becomes small enough and/or $\omega$ large enough. We propose a kinematic model of the rotary walking mode based on generic features of penetration and slip of a curved limb in granular media. The model captures the dependence of robot speed on limb frequency and the transition between walking and swimming modes but highlights the need for a deeper understanding of the physics of granular media.
\end{abstract}

Compared to agile terrestrial organisms, most man-made vehicles possess limited mobility on complex terrain \underline{\cite{kumagai04}} and are easily thwarted by materials like rubble and sand. Increased locomotive performance of engineered platforms demands better understanding of interaction with complex environments.  At the same time, there is increasing evidence that small legged machines can have greater maneuverability than large wheeled vehicles in many natural environments \underline{\cite{raibert86}}. However, while wheeled and treaded locomotion on sand has been well studied by pioneers like Bekker \underline{\cite{bekkerbook}}, study of the interaction of animals or legged devices with complex media like sand is in its infancy \underline{\cite{kadanoff99}}, in part because the physics of penetration and drag in granular media is largely unexplored for realistic conditions. Nearly all previous experiments and models of terrestrial locomotion were developed for running and walking on rigid, flat, non-slip substrates in which the possibility of substrate flow was ignored \underline{\cite{altendorferKod04,westervelt04,burridge99,berkemeier98,schaal96}}.

Rainforest, grassland, polar tundra, mountains, and desert are examples of complex Earth terrains with flowing substrates that challenge locomotors; the limited experience of the Mars Rovers supports the presumption that extraterrestrial landscapes will be even more daunting.  Deserts, common in nature and occupying about 10 percent of land surface on Earth \underline{\cite{ezcurrabook}}, consist largely of granular media, a representative complex substrate. Granular materials, defined as collections of discrete particles, can exhibit solid-like \underline{\cite{terzaghibook}} behavior below a critical yield stress \underline{\cite{neddermanbook}}, while fluid-like \underline{\cite{heil04}}, gas-like \underline{\cite{vanzon94}}, and even glass-like \underline{\cite{golAswi06}} behaviors are possible during flow. Yet, compared to other complex materials like debris, mud, or snow, granular materials are simple enough that fundamental understanding of the collective physics can be achieved through interplay of experiment and theory. Unlike more heterogeneous real-world environments, granular media can be precisely controlled using laboratory scale devices \underline{\cite{golAswi06,golAkor06}} to create states of varying material strength that mimic different deformable flowing materials produced during locomotion on complex terrains. Here we systematically explore the performance of a small legged device, SandBot, on granular media prepared in different packing states with volume fraction ranges typical of desert sand~\underline{\cite{lougeprivate}}.  Despite SandBot's (and its predecessor RHex \underline{\cite{saranli01}}) ability to move nimbly and rapidly over a wide range of natural terrain, we find that on granular media the locomotion is remarkably sensitive to substrate preparation and gait characteristics, which points to both the need for a more sophisticated understanding of the physics of motion within granular media and the possibility of better robotic design and control paradigms for locomotion on complex terrains.

\section{Results and Discussion}

The robot we study, SandBot (Fig.~\hyperlink{fig1}{\underline{1a}}), is the smallest (mass 2.3 kg) in a successful series of biologically inspired~\underline{\cite{koditschek04}} hexapedal robots, the RHex class \underline{\cite{saranli01}}. RHex incorporates the pogo-stick like dynamics observed in a diversity of biological organisms running on hard ground~\underline{\cite{bliAful}}. This dynamics, called the Spring Loaded Inverted Pendulum (SLIP) template, is hypothesized to confer passive self-stabilization properties to both biological and robotic locomotors~\underline{\cite{holAful06}}. RHex was the first legged machine to achieve autonomous locomotion at speeds exceeding one body length/s~\underline{\cite{saranli01}} and it and its ``descendants`` such as EduBot/SandBot, Whegs~\underline{\cite{ritAqui04}} and iSprawl~\underline{\cite{kimAcla06}} are still the leaders in legged mobility (roughly, speed and efficacy) on general terrain. In fact, prior to the recent development of the much larger BigDog~\underline{\cite{plaAbue06}} platform (1 m long, 75 kg), RHex remained the only class of legged machine with documented ability to navigate on complex, natural, outdoor terrain of any kind and has been used as the standard platform in comparisons with commercial wheeled and tracked vehicles like Packbot~\underline{\cite{mcbAlon03}}.

SandBot moves using an alternating tripod gait in which two sets of three approximately c-shaped legs rotate synchronously and $\pi$ out of phase. A clock signal (Fig.~\hyperlink{fig1}{\underline{1c}}), defined by three gait parameters (see Materials and Methods), prescribes the angular trajectory of each tripod. The c-legs distribute contact~\underline{\cite{spaAgol}} over their surface and allow the robot to move effectively on a variety of terrain. On rigid, no-slip ground SandBot's limb trajectories are tuned to create a bouncing locomotion~\underline{\cite{saranli01}} that generates speeds up to 2~body lengths/s ($\approx 60$~cm/s). We tested this clock signal on granular media, but found that the robot instead of bouncing adopts a swimming gait in which the legs always slip backward relative to the stationary grain bed and for which performance is reduced by a factor of 30 to $\approx 2$~cm/s (see Supporting Information Movie \href{https://www.youtube.com/watch?v=52wY6THW3To&ab_channel=TerradynamicsLab}{\underline{S1}}). We surmised that this was due to an interval of double stance (both tripods in simultaneous contact with the ground), which is useful on hard ground during bouncing gaits but apparently causes tripod interference on granular media. Changing the gait parameters to remove the double stance allowed SandBot to move (see Movie \href{https://www.youtube.com/watch?v=xiYIkDo4j30&ab_channel=TerradynamicsLab}{\underline{S2}}) in the granular media at speeds up to 1~body length/s ($\approx 30$~cm/s) in a rotary walking gait that resembles the pendular gait of the robot on hard ground~\underline{\cite{altendorfer01}} but with important kinematic differences (discussed below). No amount of gait parameter adjustment produced rapid bouncing locomotion on granular media. We hypothesize that the $\approx 50 \%$ decrease in top speed relative to hard ground is associated with the inability of the robot to undergo the arial phases associated with the bouncing gait~\underline{\cite{goldmanlizardspeed}}.

\clearpage
\manuscriptfigure{1}{height=8.05in,width=\linewidth,keepaspectratio}{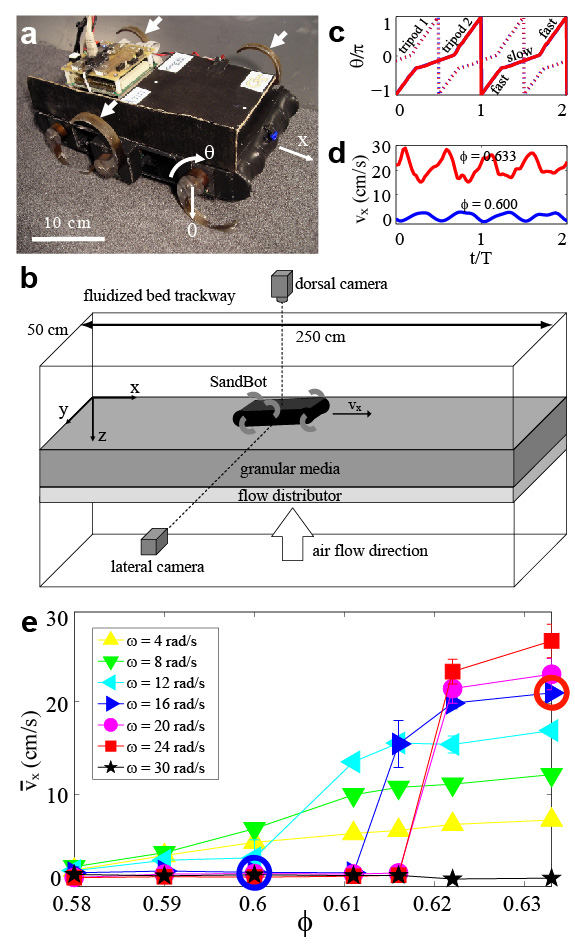}{Locomotion of a legged robot on granular media is sensitive to substrate packing and limb frequency. (a) The six-legged robot, SandBot, moves with an alternating tripod gait (alternate triplets of limbs rotate $\pi$ out of phase); arrows indicate members of one tripod. (b) Pulses of air through the bottom of the fluidized bed trackway control the initial volume fraction $\phi$ of the granular substrate; air is turned off before the robot begins to move. (c) Tripod leg-shaft angle $\theta$ vs.\ time is controlled to follow a prescribed trajectory with two phases: a slow stance phase and a fast swing phase. Overlapping trajectories from trials with $\phi=0.633$ (red) and $\phi=0.600$ (blue) at $\omega=16$~rad/s demonstrate that the controller maintains the desired kinematics independent of material state. (d) Identical tripod trajectories produce different motion for $\phi$ = 0.600 and $\phi$ = 0.633. (e) For given limb frequency ($\omega$ = 4, 8, 12, 16, 20, 24, and 30 rad/s) the robot speed is remarkably sensitive to $\phi$. Red and blue circles show the corresponding states in (c) and (d).}

In the desert, animals and man-made devices can encounter granular media ranging in volume fraction from $\phi = 0.55$ to $\phi= 0.64$~\underline{\cite{lougeprivate}}, and some desert adapted animals (like lizards) can traverse a range of granular media with little loss in performance~\underline{\cite{golAkor06}}. To test the robot performance on controlled volume fraction granular media, we employ a 2.5~m long fluidized bed trackway (Fig.~\hyperlink{fig1}{\underline{1b}}) \underline{\cite{jacksonbook}}, which allows the flow of air through a bed of granular media, in this case $\sim 1$~mm poppy seeds. With initial fluidization followed by repeated pulses of air \underline{\cite{schroter05}}, we prepare controlled volume fraction states with different penetration properties \underline{\cite{schroter07}}. In this study, we test the performance (forward speed $v_x$) of SandBot with varied leg-shaft angular frequency ($\omega$) for volume fraction ($\phi$) states ranging from loosely to closely packed ($\phi = 0.580$ to $\phi = 0.633$). We chose forward speed as a metric of performance since it could be readily measured by video imaging. We hypothesized that limb frequency would be important to robot locomotion since the substrate yield strength increases with volume fraction and the yield stress $\times$ robot limb area divided by the robot mass $\times$ velocity is proportional to the maximum limb frequency for efficient locomotion.

We find that robot performance (speed) is remarkably sensitive to $\phi$ (see Movie \href{https://www.youtube.com/watch?v=cLSPLyD69fY&ab_channel=TerradynamicsLab}{\underline{S3}}). For example, at $\omega =16$~rad/s the robot speed $v_x(t)$ shows a change in average speed $\overline{v}_x$ of nearly a factor of five as $\phi$ changes by just 5~\% (Figs.~\hyperlink{fig1}{\underline{1d}},e). For the closely packed state ($\phi= 0.633$), $\overline{v}_x \approx 20$~cm/s with $5$~cm/s oscillations during each tripod rotation, whereas for a more loosely packed state ($\phi$ = 0.600), $\overline{v}_x \approx 2$~cm/s with 1~cm/s oscillations in velocity.

This sensitivity to volume fraction is shown in the average robot speed vs.\ volume fraction (Fig.~\hyperlink{fig1}{\underline{1e}}). For fixed $\omega$, $\overline{v}_x$ is effectively constant for $\phi$ above a critical volume fraction $\phi_c(\omega)$, but is close to zero for $\phi$ below $\phi_c(\omega)$. For fixed $\omega$, $\phi_c(\omega)$ separates volume fraction into two regimes: the ``walking'' regime ($\phi \geq \phi_c$, $\overline{v}_x >>$ 0) and the ``swimming'' regime ($\phi < \phi_c(\omega)$, $\overline{v}_x \approx 2$~cm/s). See Movies \href{https://www.youtube.com/watch?v=vFs0pzXTk80&ab_channel=TerradynamicsLab}{\underline{S4}} and \href{https://www.youtube.com/watch?v=wTcyWKrKKvg&ab_channel=TerradynamicsLab}{\underline{S5}} for examples of rotary walking and swimming modes.

The rotary walking mode is dominant at low $\omega$ and high $\phi$.  In this mode, a tripod limb penetrates down and backward into the ground until the granular yield stress exceeds the limb transmitted inertial, gravitational, and frictional  stresses at a depth $d(\omega,\phi)$.  At this point, rather than rolling forward like a wheel, the c-leg abruptly stops translating relative to the grains and begins slipping tangentially in the circular depression surrounding it; at the same time, the center of rotation moves from the axle to the now stationary center of curvature  (see Fig.~\hyperlink{fig3}{\underline{3a}}).  The simultaneous halt in both vertical and horizontal leg motion is apparently due to the large reduction in friction forces which occurs when the weight of the robot is supported by the limbs rather than the underside of the body.  The ensuing rotary motion propels the axle and consequently the rest of the robot body along a circular trajectory in the $x-z$ plane with speed $R\omega$, where $R=3.55$~cm is the c-leg radius.  The forward body motion ends when, depending on $\phi$ and $\omega$, either the second tripod begins to lift the robot or the underside of the robot contacts the ground.

With increased $\omega$ limbs penetrate further as the requisite force to rapidly accelerate the robot body to the finite limb speed ($R\omega$) increases.  As the penetration depth approaches its maximum $2R-h$, where $h=2.5$~cm is the height of the axle above the flat underside of the robot, the walking step size goes to zero since there is no longer a point in the cycle where the limb ceases its motion relative to the grain bed.  Any subsequent forward motion is due solely to thrust forces generated by the swimming-like relative translational motion of the limb though the grains. Note that $\phi_c(\omega)$ increases with $\omega$, and that the transition from rotary walking to swimming is sharper in $\overline{v}_x$ for higher $\omega$ and smoother for lower $\omega$. The much slower swimming mode occurs for all volume fractions for $\omega \geq 28$~rad/s.

\manuscriptfigure{2}{width=4.6in}{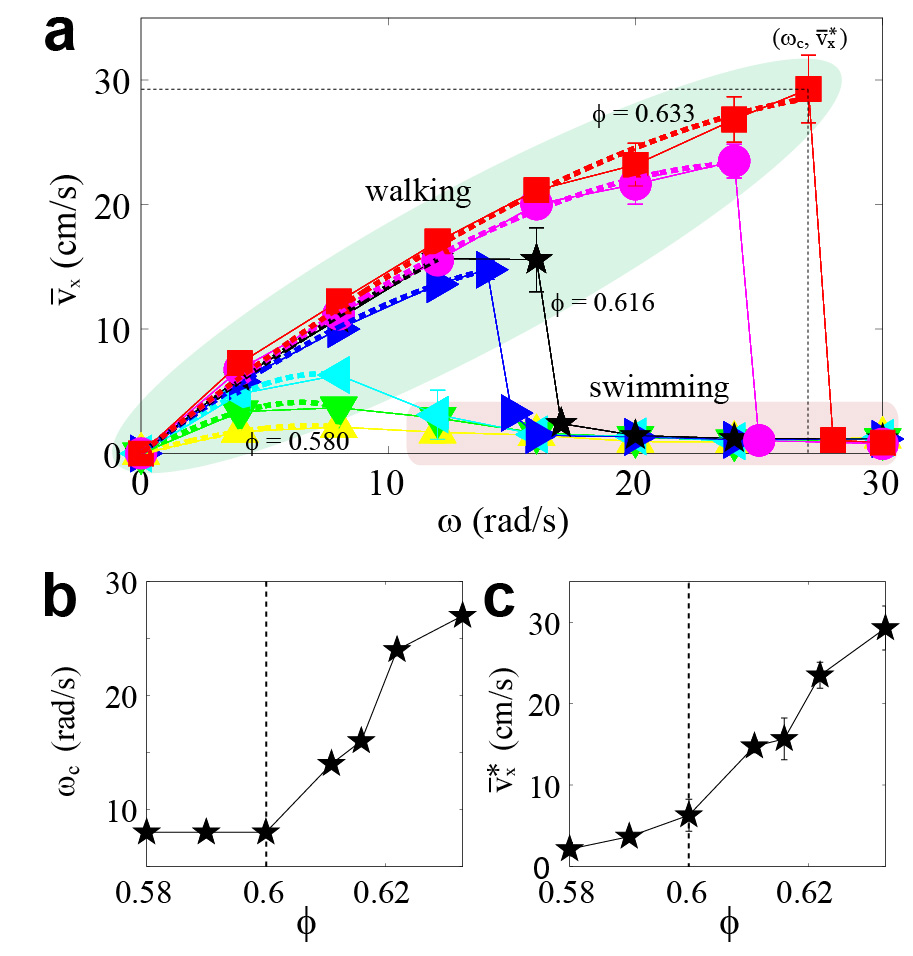}{Average robot speed vs.\ limb frequency. (a) For a given volume fraction $\phi$, $\overline{v}_x$ increases sub-linearly with $\omega$ to a maximal average speed $\overline{v}^{*}_x$ at a critical limb frequency $\omega_c$ above which the robot swims ($\overline{v}_x \sim 2$~cm/s). The solid lines and symbols are for $\phi = 0.580$, 0.590, 0.600, 0.611, 0.616, 0.622, and 0.633. The dashed lines are fits from a simplified model discussed in the text. (b)(c) The dependence of $\omega_c$ and $\overline{v}^{*}_x$ on $\phi$ shows transitions at $\phi \approx$ 0.6 (dashed lines).}

Plotting the average robot speed as a function of limb frequency (Fig.~\hyperlink{fig2}{\underline{2a}}) shows how the robot suffers performance loss as its legs rotate more rapidly. For fixed $\phi$, $\overline{v}_x$ increases sub-linearly with $\omega$ to a maximal speed $\overline{v}^{*}_x$ at a critical limb frequency $\omega_c$, above which $\overline{v}_x$ quickly decreases to $\approx 2$~cm/s (swimming)~\underline{\cite{resolutionnote}}. The robot suffers partial performance loss (mixed walking and swimming) below $\omega_c$ and near total performance loss (pure swimming) above $\omega_c$. Performance loss for $\phi \gtrapprox 0.6$ is more sudden ($\Delta \omega \approx 1$~rad/s) compared to performance loss for $\phi \lessapprox 0.6$. Both $\omega_c$ and $\overline{v}^{*}_x$ display transitions at $\phi \approx 0.6$ (Figs.~\hyperlink{fig2}{\underline{2b}},c). The transition at $\phi \approx 0.6$ for the rapidly running robot is noteworthy since it has been observed that granular media undergo a transition in quasi-static penetration properties at $\phi \approx 0.6$~\underline{\cite{schroter07}}.

Starting with the observed kinematics of rotary walking with circular slipping, we constructed a straightforward two-parameter model that captures the essential elements determining granular locomotion for our legged device and agrees well with the data (dashed lines in Fig.~\hyperlink{fig2}{\underline{2a}}).  The model, which incorporates simplified kinematics and granular penetration forces while still agreeing well with a more realistic treatment (for a more detailed discussion of the model, see Materials and Methods), indicates that reduction of step length through increased penetration depth is the cause of the sub-linear increase in $\overline{v}_x$ with $\omega$ and the rapid loss of performance above $\omega_c$.  The model assumes that the two tripods act independently, that the motion of each tripod can be understood by examining the motion of a single c-leg supporting a mass $m$ equal to one-third of SandBot's total mass, and that the underside of the robot rests on the surface at the beginning of limb/ground contact.

\manuscriptfigure{3}{width=6in}{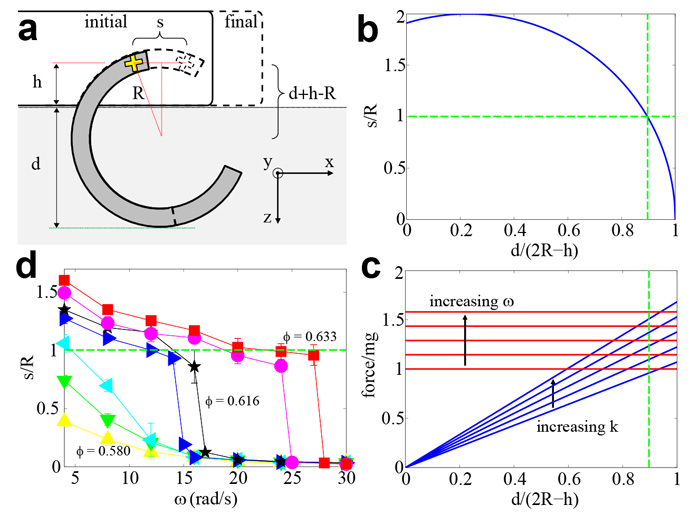}{(a) Schematic of a single robot leg during a step in granular media. After reaching penetration depth $d$, the leg rotates about its center and propels the robot forward a step length $s$. The solid shape denotes the initial stage of the rotational motion and the dashed shape indicates when the limb begins to withdraw from the material (end of forward body motion).  (b) Step length vs.\ penetration depth (blue) with critical step size (green dashed horizontal) and critical penetration depth (green dashed vertical) indicating where the robot begins to encounter ground disturbed by the previous step.  (c) Granular penetration force for $k=1.75,2.00,2.25,2.50,2.75\times 10^5$~N/m (blue) and force required to initiate rotary walking for $\omega=0,8,16,24,32$~rad/s (red) vs.\ penetration depth using simplified walking model with $\Delta t=0.2$~s.  The penetration depth at constant $\phi$ is determined by the intersections of the corresponding blue line with the red lines. Beyond the critical depth (green dashed line) limbs encounter disturbed material and move to lower blue lines.  (d) Step length as a function of $\omega$ derived from $2s=2\pi\overline{v}/\omega$ reveals the condition for the onset of swimming as $s/R \approx 1$.}

Using the geometry of rotary walking (see Fig.~\hyperlink{fig3}{\underline{3a}}), the walking step size per c-leg rotation is \[s=2\sqrt{R^2-(d+h-R)^2},\] where $d$ is the maximum depth of the lowest point on the leg.  After the robot has advanced a distance $s$, the body again contacts the ground and the c-leg moves upward.  Since during each clock signal period there are two leg rotations (one for each tripod), the average horizontal velocity is $2s$ $\times$ limb frequency or $\overline{v}_x= \frac{\omega s}{\pi}.$  The maximum limb penetration depth $d$ is thus the key model component as it controls the step length (see Fig.~\hyperlink{fig3}{\underline{3b}}) and consequently the speed.  Maximum limb penetration depth is determined by balancing the vertical acceleration of the robot center of mass $ma$ with the sum of the vertical granular penetration force $kz$~\underline{\cite{albert99}} and the gravitational force $mg,$ where $g$ is the acceleration due to gravity, and $k(\phi)$ is a constant characterizing the penetration resistance of the granular material of volume fraction $\phi$.

At small $\omega$, $ma \approx 0 = \sum F_i = mg - kd$ so $d = mg/k$, which is the minimum penetration depth. For finite $\omega$ the penetration depth is greater since an additional force must be supplied by the ground to accelerate the robot body to the leg speed $R\omega$ when the c-leg stops translating in the material.  Taking $a=\Delta v/\Delta t$, with $\Delta v = R\omega-0$ and $\Delta t$ the characteristic elastic response time of the limb and grain bed, gives the acceleration magnitude $a=R\omega/\Delta t.$  The direction of the acceleration depends on the position of the c-leg.  To keep the model simple we approximate the vertical component of the acceleration with its magnitude. Equating the vertical forces with mass $\times$ acceleration (see Fig.~\hyperlink{fig3}{\underline{3c}}), $ -m\frac{R\omega}{\Delta t} = mg - kd$, gives c-leg penetration $d=\frac{m}{k}(\frac{R\omega}{\Delta t}+g)$ with average horizontal velocity $\overline{v}_x= \frac{2R\omega}{\pi}\sqrt{1-\left[\frac{m}{k(\phi)}(\frac{\omega}{\Delta t} + \frac{g}{R})+\frac{h}{R}-1\right]^2}$.  Fits to this model are indicated by dashed lines in Fig.~\hyperlink{fig2}{\underline{2b}}.  The expression captures the sub-linear increase in $\overline{v}_x$ with $\omega$ at fixed $k(\phi)$, the increase in speed at fixed $\omega$ as the material strengthens (increasing $k$ with increasing $\phi$), and transition to zero rotary walking velocity when $\omega$ is sufficiently large.

The expression for $\overline{v}_x$ is determined by the two fit parameters $k$ and $\Delta t$.  The parameter $k$ characterizing the penetration resistance increases monotonically with $\phi$ from 170 to 220~N/m and varies rapidly below $\phi \approx 0.6$ and less rapidly above.  Its average value of $\approx 200$~N/m corresponds to a shear stress per unit depth of $\alpha \approx 470$~kN/m$^3$ (using leg area=$w R$ where $w$ is the leg width) which is in good agreement with penetration experiments we performed on poppy seeds that yield $\alpha =300$ and 480~kN/m$^3$ for $\phi=0.580$ and 0.622 respectively and is comparable to previous measurements of slow penetration into glass beads~\underline{\cite{stoAbar04}} where $\alpha \approx 250$~kN/m$^3.$   In contrast, $\Delta t$ varies little with $\phi$ and has an average value of 0.4~s compared to the robot's measured hard ground oscillation period of 0.2~s when supported on a single tripod.  The differences in $\Delta t$ can be understood as follows.  In our model we assume the two tripods do not simultaneously contact the ground; however, in soft ground this is not the case, which consequently reduces the effective step size per period from $2s$ to a lesser value.  The fit value of $\Delta t$ is sensitive to this variation; reducing the step size (and thus the speed) in the experimental data by just 13~\% decreases $\Delta t$ to 0.2~s while $k$ is increased by less than 10~\%.

Our model indicates that for deep penetration the walking step length is sensitive to penetration depth ({\em e.g.\ } Fig. \hyperlink{fig3}{\underline{3b}}).  As the walking step size goes to zero with increasing $\omega$ or decreasing $\phi$, the fraction of the ground contact time that the leg slips through the grains (swimming) goes to one. Swimming in granular media differs from swimming in simple fluids as the friction dominated thrust and drag forces are largely rate independent at slower speeds~\underline{\cite{albert99,wieghardt75}}.  When thrust exceeds drag and using constant acceleration kinematics, the robot advances a distance proportional to the net force divided by $\omega^2$ per leg rotation, and, consequently, speed is proportional to $\omega^{-1}$. This explains the weak dependence of $\overline{v}_x$ on $\omega$ in the swimming mode.  The increase in robot speed with decreasing $\omega$ is bounded by the condition that the robot speed in a reference frame at rest with respect to the ground cannot exceed the horizontal leg speed in a reference frame at rest with respect to the robot's center of mass.  This condition ensures the existence of and eventual transition to a walking mode as $\omega$ is decreased.

The transition from walking to swimming appears gradual for $\phi \lessapprox 0.6$ since the penetration depth increases slowly with $\omega$ at small $\omega$ ($R\omega/\Delta t \ll g$) and the $\omega^{-2}$ contribution to the per cycle displacement from swimming is relatively large (see {\em e.g.\ } the data at $\omega=12$~rad/s in Fig.~\hyperlink{fig3}{\underline{3b}}).  However, for $\phi \gtrapprox 0.6$, the transition is abrupt. This sharp transition occurs because the step size is reduced sufficiently that the legs encounter material disturbed by the previous step; we hypothesize that the disturbed material has lower $\phi$ and $k$.  At higher $\phi$, the volume fraction of the disturbed ground is significantly less than the bulk which increases penetration and consequently greatly reduces $s$.  This is not the case for the transition from walking to swimming at lower $\phi$ (and low $\omega$) where the volume fraction of the disturbed material is largely unchanged relative to its initial value.  For the robot to avoid disturbed ground it must advance a distance $R$ on each step, {\em i.e.\ } $s \geq R$, or in terms of the penetration depth, $d \leq (\frac{\sqrt{3}}{2}+1)R-h = 5.0$~cm (green dashed lines in Fig.~\hyperlink{fig3}{\underline{3b}},c).  The disturbed ground hypothesis is supported by calculations of the step size derived from the average velocity $2s=2\pi\overline{v}_x/\omega$ which show a critical step size near $s/R=1$ at the walking/swimming transition (Fig.~\hyperlink{fig3}{\underline{3d}}). The somewhat smaller value of $s/R \approx 0.9$ evident in the figure can be understood by recognizing that for $s$ slightly smaller than $R$ the majority of the c-leg still encounters undisturbed material.  Signatures of the walking/swimming transition are also evident in lateral views of the robot kinematics (see Movies \href{https://www.youtube.com/watch?v=cLSPLyD69fY&ab_channel=TerradynamicsLab}{\underline{S3}}, \href{https://www.youtube.com/watch?v=vFs0pzXTk80&ab_channel=TerradynamicsLab}{\underline{S4}}, and \href{https://www.youtube.com/watch?v=wTcyWKrKKvg&ab_channel=TerradynamicsLab}{\underline{S5}}).

At higher $\omega$ in the swimming mode, limbs moves with sufficient speed to fling material out of their path and form a depression which reduces thrust because the limbs are not as deeply immersed on subsequent passes through the material. However, as limb speed increases further, thrust forces becomes rate dependent and increase because the inertia imparted to the displaced grains is proportional to $\omega^2$.  Between strokes, the excavated depression refills at a rate dependent on the difference between the local surface angle and the angle of repose~\underline{\cite{devAdeb07}} and the depression size.  Investigating the competition between these different processes at high $\omega$ and their consequences for locomotion could be relevant to understanding how to avoid becoming stranded or to free a stranded device.

\section{Conclusions}

Our studies are the first to systematically investigate the performance of a legged robot on granular media, varying both properties of the medium (volume fraction) and properties of the robot (limb frequency). Our experiments reveal how precarious it can be to move on such complex material: changes in $\phi$ of less than one percent result in either rapid motion or failure to move, and slight kinematic changes have a similar effect. A kinematic model captures the speed dependence of SandBot on granular material as a function of $\phi$ and $\omega$. The model reveals that the sublinear dependence of speed on $\omega$ and the rapid failure for sufficiently small $\phi$ and/or large $\omega$ are consequences of increasing limb penetration with decreasing $\phi$ and/or increasing $\omega$, and changes to local $\phi$ due to penetration and removal of limbs. While detailed studies of impact and penetration of simple rigid objects exist~\underline{\cite{golAumb08,albert99}}, further advances in performance (including increases in efficiency) and design of limb geometry will require a more detailed understanding of the physics associated with penetration, drag, and crater formation and collapse, especially their dependence on $\phi$. Better understanding of this physics can guide development of theory of interaction with complex media advanced enough to predict limb design~\underline{\cite{sanAspe07}} and control~\underline{\cite{raibert91}} strategies, similar to the well-developed models of aerial and aquatic craft. Analysis of physical models such as SandBot can also inform locomotion biology in understanding how animals appear to move effortlessly across a diversity of complex substrates~\underline{\cite{golAche06,spaAgol}}. Such devices will begin to have capabilities comparable to organisms; these capabilities could be used for more efficient and capable exploration of challenging terrestrial ({\em e.g.\ } rubble and disasters sites) and extra-terrestrial ({\em e.g.\ } the Moon and Mars) environments.

\section{Materials and Methods}

{\em Limb kinematics} SandBot's six motors are controlled by a clock signal to follow the same prescribed kinematic path during each rotation and, as shown in previous work on RHex, changes in these kinematics have substantial effects on robot locomotor performance~\underline{\cite{WeiAGro04}}. The controlling clock signal consists of a fast phase and a slow phase with respective angular frequencies. The fast phase corresponds to the swing phase, and the slow phase corresponds to the stance phase. A set of three gait parameters uniquely determines the clock signal configuration: $\theta_s$, the angular span of the slow phase; $\theta_0$, the leg-shaft angle of the center of the slow phase; and $d_c$, the duty cycle of the slow phase. Specifying the cycle average limb angular frequency $\omega$ fully determines the limb motion.

In pilot experiments we tested two sets of clock signals: a hard ground clock signal (HGCS) with ($\theta_s$ = 0.85 rad, $\theta_0$ = 0.13 rad, $d_c$ = 0.56) which generates a fast bouncing gait (60~cm/s) on hard ground \underline{\cite{saranli01}} but very slow ($\sim 1$~cm/s) motion on granular media, and a soft ground clock signal (SGCS) with ($\theta_s$ = 1.1 rad, $\theta_0$ = -0.5 rad, $d_c$ = 0.45) which produces unstable motion on hard ground but regular motion on granular media. These experiments showed that the locomotor capacity of SandBot is sensitive to the clock signal. Careful observation of limb kinematics revealed that the hard ground clock signal fails on granular media because of the simultaneous stance phase of two tripods. In this study, we use SGCS and explore robot performance as a function of limb frequency and substrate volume fraction.

Integrated motor encoders record the position and current (and thus torque) of SandBot's motors vs.\ time. Comparison of the measured and prescribed angular trajectories for both sets of timing parameters show a high degree of fidelity with an error of a few percent.  Therefore, SandBot's change in performance between HGCS and SGCS timing comes from the physics of the substrate interaction.

{\em Trackway volume fraction control} To systematically test SandBot's performance vs.\ substrate volume fraction, we employ a 2.5~m long, 0.5~m wide fluidized bed trackway with a porous plastic (Porex) flow distributor (thickness 0.64 cm, average pore size 90 $\mu m$). Four 300 L/min leaf blowers (Toro) provide the requisite air flow. Poppy seeds are chosen as the granular media because they are similar in size to natural sand \underline{\cite{bagnold}} and are of low enough density to be fluidized. The air flow across the fluidized bed is measured with an anemometer (Omega Engineering FMA-900-V) and is uniform to within 10 percent.

A computer controlled fluidization protocol sets the volume fraction and thus the mechanical properties of the granular media. A continuous air flow initially fluidizes the granular media in the bubbling regime. The flow is slowly turned off leaving the granular media in a loosely packed state ($\phi$ = 0.580). Short air pulses (ON/OFF time = 0.1/1~s) pack the material~\underline{\cite{schroter05}}. Increasing the number of pulses increases $\phi$ up to a maximum of $\phi = 0.633$. Volume fraction is calculated by dividing the total grain mass by the bed volume and the intrinsic poppy seed density. The mass is measured with a precision scale (Setra). The density of the granular media is measured by means of displacement in water. In experiment, since the horizontal area of the fluidized bed trackway is fixed, volume fraction is set by controlling the height of the granular media ({\em e.g.\ } volume = area $\times$ height).

{\em Kinematics measurements} To characterize SandBot's motion, we record simultaneous dorsal and lateral views with synchronized high speed video cameras (AOS Switzerland) at 100~frames/s. The center of mass (dorsal landmark) and the axles of the right-side front and rear motors (lateral landmarks) are marked with reflective material (WhiteOut). A rail-pulley system allows the robot's power and communication cables to follow the robot as it moves to minimize the drag from the cables. For each trial, we prepare the trackway with the desired volume fraction and place the robot on the prepared granular media at the far end of the trackway with both tripods in the same standing position. An LED on the robot synchronizes the video and robot motor encoder data.  After each trial, MATLAB (The MathWorks) is used to obtain landmark coordinates from the video frames and calculate $v_x(t)$. Three trials were run for each combination of ($\phi$, $\omega$) that was tested.

{\em  Detailed discussion of rotary walking locomotion model} The model presented in the main body of the manuscript simplifies the underlying physics while capturing the essential features determining robot speed. Here we describe a more complete model (which lacks a simple expression for $\overline{v}_x$) and compare its predictions to those of the simple model.  The exact expression for the vertical acceleration component of the body when the limbs gain purchase is $ma_z = ma \sin{\theta} = ma \sqrt{\frac{2(h+z)}{R} - \left (\frac{h+z}{R} \right)^2}$ instead of the approximation $ma_z = ma$ used in the simple model.  Using the exact expression, the vertical granular force necessary for walking still has the same peak value of $m(a+g)$ but decreases to $mg$ when the limb is at its lowest point.

\manuscriptfigure{4}{width=6in}{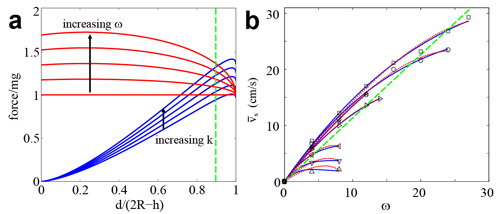}{(a) Non-dimensionalized granular force (blue curves) for $\alpha \times 10^{-4}=2.50,$ $2.75,$ $3.00,$ $3.25,$ and $3.50$~g~cm$^{-2}$s$^{-2}$ and the required force to initiate rotary walking, $a_z/g+1$ (red curves) for $\omega=0,$ 5, 10, 15, 20, 25 and 30~rad/s for the full model as a function of limb penetration depth with a $225^{\circ}$ c-leg arc angle and $\Delta t=0.15$~s.  The intersection of the red and blue curves determines the penetration depth of the limb and consequently the step size.  At constant material strength (blue) $d$ increases with increasing $\omega$, while at constant $\omega$ increasing material strength reduces $d$.  The vertical green dashed line indicates the critical penetration depth beyond which the leg encounters material disturbed by the previous step.  (b) Comparison of $\overline{v}_x$ vs.\ $\omega$ for simple (red dotted curve) and full (blue solid curve) models.  Models are fit to the measured robot speed (symbols) for $\overline{v}_x \leq \overline{v}^*_x$.  The green dashed line indicates $\overline{v}^*_x = R\omega/\pi$ or equivalently $s=R$. In both figures $h=2.5$~cm, $R=3.55$~cm, $w=1.2$~cm and $m=767$~g.}

The second approximation we employed in the simple model is that the grain force on the leg is $kz$.  This expression is only strictly valid for a flat bottomed vertically penetrating intruder~\underline{\cite{albert99}}.  Since the leg is a circular arc, the leg-grain contact area and the vertical component of the grain force are functions of limb depth and leg-shaft angle.  Generalizing $kz$ to a local isotropic yield stress given by $\alpha z$~\underline{\cite{neddermanbook}}, the vertical force on a small segment of the limb $R d\psi$ in length at depth $z$ is $dF_z = w \alpha z R d \psi \cos \psi,$ where $w$ is the limb width and $\psi$ the angular position of the segment with respect to a vertical line passing through the axle.  The total vertical component of the force acting on the leg is then $R w \alpha \int_{-\psi_{min}}^{\psi_{max}} z \cos \psi d\psi.$ Substituting $z(\psi) = R (\cos \psi -1) + d$ and integrating gives $F_z = R w \alpha \left [ \frac{R}{2} (\psi + \cos \psi \sin \psi) + (d-R) \sin{\psi} \right ]_{-\psi_{min}}^{\psi_{max}},$ where $\psi_{max} = \cos^{-1} (1-d/R)$ and $\psi_{min}=\psi_{max}$ when the leg tip is above the center of the c-leg and $\cos^{-1} \left ( \frac{d+h}{R} -1 \right) + \Delta \xi$ when it is below the center of the c-leg. $\Delta \xi$ is the angular extent of the limb beyond $\pi$ ({\em e.g}~$\Delta \xi=0$ for a semi-circular limb).

Fig.~\hyperlink{fig4}{\underline{4a}} shows that the full model using realistic parameters shares the same essential physics as the simple model.  For a given material strength (blue curves), the penetration depth increases with increasing $\omega$ (intersection of blue and red curves) until either the step size is reduced below the critical value (vertical green dashed line) or the granular force no longer provides the required force to support the robot's mass and accelerate it.  Fig.~\hyperlink{fig4}{\underline{4b}} presents fits to the experimental data of the average speed $\overline{v}_x$ vs.\ $\omega$ for the full and simple models for $\overline{v}_x \leq \overline{v}^*_x$ at each $\phi.$ The fits and fit parameters for the simple ($\overline{\Delta t}=0.4$~s, $\overline{\alpha}=470$~kN/m$^3$) and full ($\overline{\Delta t}=0.2$~s, $\overline{\alpha}=330$~kN/m$^3$) models are in good agreement when the step size is less than the critical value $s=R.$

\section{Acknowledgments}

We thank Daniel Cohen, Andrew Slatton and Adam Kamor for helpful discussion. This work was supported by the Burroughs Wellcome Fund (D.I.G., C.L., P.B.U).

\section*{Supporting movies}
{\fontsize{10}{13}\selectfont\begin{enumerate}\setlength{\itemsep}{3pt}\setlength{\parskip}{0pt}
\item \href{https://www.youtube.com/watch?v=VFx866294XY&ab_channel=TerradynamicsLab}{\underline{Control of Granular Media Compaction by Air Fluidized Bed}}
\item \href{https://www.youtube.com/watch?v=W9DlAiHQ_bg&ab_channel=TerradynamicsLab}{\underline{SandBot Gaits}}
\item \href{https://www.youtube.com/watch?v=52wY6THW3To&ab_channel=TerradynamicsLab}{\underline{Movie S1: SandBot Stuck with Gait Tuned for Solid Ground}}
\item \href{https://www.youtube.com/watch?v=xiYIkDo4j30&ab_channel=TerradynamicsLab}{\underline{Movie S2: SandBot Moving with Gait Tuned for Granular Media}}
\item \href{https://www.youtube.com/watch?v=cLSPLyD69fY&ab_channel=TerradynamicsLab}{\underline{Movie S3: SandBot on Granular Media of Different Compactions}}
\item \href{https://www.youtube.com/watch?v=vFs0pzXTk80&ab_channel=TerradynamicsLab}{\underline{Movie S4: SandBot Walking Mode}}
\item \href{https://www.youtube.com/watch?v=wTcyWKrKKvg&ab_channel=TerradynamicsLab}{\underline{Movie S5: SandBot Swimming Mode}}
\end{enumerate}}
\end{document}